\documentclass[sigconf,nonacm]{acmart}
\usepackage{multirow, booktabs}
\usepackage{listings}

\AtBeginDocument{%
  \providecommand\BibTeX{{%
    \normalfont B\kern-0.5em{\scshape i\kern-0.25em b}\kern-0.8em\TeX}}}

\setcopyright{acmcopyright}
\copyrightyear{2026}
\acmYear{2026}
\acmDOI{XXXXXXX.XXXXXXX}

\acmConference[Conference'26]{ACM Conference}{June 08--12, 2026}{Somewhere, Earth}
\acmPrice{15.00}
\acmISBN{978-1-4503-XXXX-X/26/06}

\begin{document}

\title{Evaluating Gemma4 Models as AI Teaching Assistants for Introductory Parallel Programming: A DataRaceBench Study}

\author{Sabbir Hussain Meraj}
\affiliation{%
  \institution{The University of Texas at San Antonio}
  \city{San Antonio}
  \country{USA}}
\email{sabbirhussain.meraj@utsa.edu}

\author{Riham Chowdhury}
\affiliation{%
  \institution{The University of Texas at San Antonio}
  \city{San Antonio}
  \country{USA}}
\email{riham.chowdhury@utsa.edu}

\author{Shimul Debnath}
\affiliation{%
  \institution{The University of Texas at San Antonio}
  \city{San Antonio}
  \country{USA}}
\email{shimul.debnath@utsa.edu}

\author{Wei Wang}
\affiliation{%
  \institution{The University of Texas at San Antonio}
  \city{San Antonio}
  \country{USA}}
\email{wei.wang@utsa.edu}

\renewcommand{\shortauthors}{Meraj et al.}

\begin{abstract}
Debugging data races is a major challenge for students learning parallel programming due to the non-deterministic nature of concurrent execution and the complexity of shared-memory semantics. Recent advances in Large Language Models (LLMs) suggest that they could serve as AI teaching assistants, but the capabilities of lower-cost open-weight models for parallel debugging remain unclear. In this paper, we evaluate two Gemma4 open-weight models, Gemma4-E4B and Gemma4-31B, on their ability to identify, explain, and repair data races in OpenMP programs from the DataRaceBench benchmark suite. We also investigate whether contextual hints, including ThreadSanitizer (TSan) reports and model-generated explanations, improve repair quality. Our results show that Gemma4-E4B correctly explained 82 of 104 race-condition programs and successfully repaired 73, while Gemma4-31B achieved 100 correct explanations and 98 successful repairs. Surprisingly, additional context did not consistently improve repair effectiveness and sometimes reduced performance. These findings suggest that open-weight LLMs can provide valuable support for student self-debugging, with larger models offering near-complete coverage of the benchmark suite.
\end{abstract}


\maketitle

\section{Introduction}
Undergraduate students learning parallel programming frequently struggle with the complexities of debugging multi-threaded code~\cite{ParallelEduDebug,ParaVis}. The non-deterministic nature of data races, combined with the subtle semantics of shared memory models, creates a steep learning curve for novices. Supporting students during this process is equally challenging for instructors, who often have limited time to provide the personalized, line-by-line debugging assistance required in large introductory courses.

Large Language Models (LLMs) have recently demonstrated significant capabilities in code analysis and debugging, suggesting they could serve as effective AI teaching assistants \cite{xia2023llmapr,wei2023copiloting,xia2024chatrepair}. By providing immediate feedback and explanations, LLMs could potentially empower students to debug their parallel programs independently. However, high-performance commercial LLM models are often expensive to students and universities. It is therefore worthwhile to explore whether open-weight LLM models, which can be deployed locally and at a lower cost, are sufficiently capable of supporting parallel debugging tasks.

In this paper, we evaluate the performance of two Gemma4 open-weight models \cite{gemmateam2024gemma} in identifying, explaining, and fixing data races. Our investigation is guided by three primary research questions:
\begin{itemize}
    \item \textbf{RQ1: Identification and Explanation.} To what extent can the Gemma4 models identify and explain the underlying causes of data races in OpenMP programs?
    \item \textbf{RQ2: Fix Generation Capability.} Can the Gemma4 models generate correct and parallel-preserving fixes for data races across various OpenMP constructs?
    \item \textbf{RQ3: Impact of Contextual Hints.} Does the inclusion of runtime tool output (ThreadSanitizer) and model-generated explanations improve the quality and correctness of the generated fixes?
\end{itemize}

We conduct our evaluation using DataRaceBench, a benchmark suite of C/C++ programs with known data races \cite{liao2017dataracebench,chen2023dataracebench}. Our experimental setup involves a multi-step pipeline where models generate explanations and fixes based on various levels of context, ranging from raw code to code augmented with TSan reports \cite{serebryany2009threadsanitizer}. We evaluate two scales of the Gemma model \cite{gemmateam2024gemma}: the small-scale Gemma4-E4B and the larger Gemma4-31B, running on NVIDIA H200 GPUs. The correctness of the generated fixes is verified through a combination of runtime analysis with TSan and a multi-model audit involving Gemini 3.1 \cite{google2026geminimodels}, Claude Opus 4.7 \cite{anthropic2026claudemodels}, and GPT 5.5 \cite{openai2026models}, as well as human manual evaluation.

Our results show that Gemma4-E4B correctly explained 82 out of 104 race-condition programs and successfully repaired 73 programs. These findings suggest that the model has the potential to assist students in understanding and debugging race conditions.
We also find that the larger Gemma4-31B model offers significantly better performance (with 100 explained and 98 fixed) and should be used when resources allow. Surprisingly, we observe that adding more context, such as TSan outputs, does not necessarily improve fix correctness and can sometimes lead to contextual distraction, diverging from prior research results~\cite{2024-Chen-ICLR-SelfDebug}.

The primary contributions of this work are:
\begin{enumerate}
    \item A thorough evaluation of open-weight models on their ability to explain complex OpenMP data race conditions.
    \item A comprehensive analysis of race condition remediation using open-weight models across different levels of technical context and hints.
\end{enumerate}


\section{Evaluation Setup}
In this section, we describe the environment, tools, and the multi-step methodology used to evaluate Gemma's capability in diagnosing and fixing data races.

\subsection{Environment}
All experiments were conducted using \texttt{llama.cpp} with Gemma4 models. We evaluated two model scales: Gemma4-E4B (quantized to 8-bit) and Gemma4-31B (quantized to 16-bit). The inference was performed on a server equipped with NVIDIA H200 GPUs. We followed Google's default parameters, i.e., temperature is 1.0, top\_p is 0.95, and top\_k is 64. Context size is set to 262144 for both models, although Gemma4-E4B can effectively use only 128K context.

\subsection{Methodology}
Our evaluation follows a five-step pipeline designed to mimic the workflow of a student using an AI assistant to debug parallel code.

\textbf{Step 1: Explanation Generation.}
We provided the source code of all 104 C/C++ programs with race conditions from DataRaceBench to the model (with all comments removed to prevent leaking the ground truth). The model is tasked with identifying and explaining all data races. The prompt used for this step is shown in Listing \ref{lst:prompt_explain}.

\begin{lstlisting}[basicstyle=\ttfamily\scriptsize, frame=single, float, caption={Explanation prompt.}, label={lst:prompt_explain}]
### Task:
You are an expert in parallel programming and data race detection.
Analyze the following code and explain ALL data races present.
Instructions:
1. Start directly with "Data Race 1:" - no preamble
2. No introductory text like "Here is the analysis" or "I found"
3. No concluding remarks or summaries
4. Only the data race analysis

Format:
Data Race 1:
- Variable: [name]
- Lines: [numbers]
- Type: [read-write / write-write]
- Reason: [explanation]

Code:
{code}

Analysis:
\end{lstlisting}

\textbf{Step 2: TSan Report Generation.}
To provide the model with technical runtime feedback, we execute the original DataRaceBench programs using ThreadSanitizer (TSan). We utilize \texttt{libarcher}~\cite{atzeni2016archer} to improve TSan's support for OpenMP and reduce false positives. The resulting TSan reports are captured for use in later steps.

\textbf{Step 3: Basic/Direct Fix Generation.}
As the basic code fix attempt, we provide the comment-free source code to the model and request a direct fix. The prompt (Listing \ref{lst:prompt_basic}) explicitly instructs the model to correct race condition while preserving parallelism.

\begin{lstlisting}[basicstyle=\ttfamily\scriptsize, frame=single, float, caption={Basic fix prompt.}, label={lst:prompt_basic}]
### Task:
You are a code analysis assistant, expert in parallel programming. 
Correct the following code to remove the data race.
Instructions:
- Your response must be only contain the corrected code.
- The corrected code must be in the same language as the original 
  code.
- The corrected code must be free of data races.
- The corrected code must be free of undefined behavior.
- Do NOT remove OpenMP pragmas or parallel regions.
- Do NOT provide any explanations, reasoning, or extra text.

Code:
{code}
\end{lstlisting}

\textbf{Step 4: Context-Aware Fix Generation.}
To evaluate the impact of external hints, we provide the model with the same code along with its own explanation (from Step 1) and/or the TSan report (from Step 2). These "Context-Aware" prompts are shown in Listing~\ref{lst:prompt_context_explain},  Listing~\ref{lst:prompt_context_tsan}, and Listing~\ref{lst:prompt_context_explain_tsan}.

\begin{lstlisting}[basicstyle=\ttfamily\scriptsize, frame=single, float, caption={Fix with explanation prompt.}, label={lst:prompt_context_explain}]
### Task:
You are an expert in parallel programming. Fix the data races in 
the code based on the analysis provided.
Instructions:
1. Output ONLY the corrected code - no explanations
2. Do NOT remove OpenMP pragmas or parallel regions
3. Add proper synchronization based on the data race analysis
4. Preserve all original functionality and parallelism
5. Start directly with #include or the first line of code
6. No markdown, no comments, no extra text

Code:
{code}

Data Race Analysis:
{explanation}

Corrected Code:
\end{lstlisting}

\begin{lstlisting}[basicstyle=\ttfamily\scriptsize, frame=single, float, caption={Fix with TSan prompt.}, label={lst:prompt_context_tsan}]
### Task:
You are an expert in parallel programming and thread safety. 
Fix the data races in the code using the ThreadSanitizer (TSan) 
report provided.
    
Instructions:
1. Output ONLY the corrected code - no explanations
2. Do NOT remove or weaken any OpenMP pragmas or parallel regions
3. Fix ONLY what TSan reported - do not over-synchronize
4. Apply the minimal fix in this order of preference:
  - OpenMP: `reduction`, `atomic`, `#pragma omp critical`
  - C++ std: `std::atomic` for simple variables, `std::mutex` 
    for complex state
5. Preserve all original functionality and output behavior
6. Start directly with #include or the first line of code
7. No markdown, no comments, no extra text

Code:
{code}

TSan Report:
{tsan_output}

Corrected Code:
\end{lstlisting}

\textbf{Step 5: Verification.}
The correctness of the generated fixes is verified through a rigorous process. First, the fixed code is executed under TSan to check for any remaining or newly introduced data races. Second, the fixes are audited by three frontier models: Gemini 3.1, Claude Opus 4.7, and GPT 5.5. In cases where the models disagree, a manual review is conducted.

\begin{lstlisting}[basicstyle=\ttfamily\scriptsize, frame=single, float, caption={Fix with race condition explanation and TSan prompt.}, label={lst:prompt_context_explain_tsan}]
### Task:
You are an expert in parallel programming and thread safety. 
Fix the data races in the code using the ThreadSanitizer (TSan) 
report provided.
    
Instructions:
1. Output ONLY the corrected code - no explanations
2. Do NOT remove or weaken any OpenMP pragmas or parallel 
regions
3. Fix ONLY what TSan reported - do not over-synchronize
4. Apply the minimal fix in this order of preference:
   - OpenMP: `reduction`, `atomic`, `#pragma omp critical`
   - C++ std: `std::atomic` for simple variables, `std::mutex` 
   for complex state
5. Preserve all original functionality and output behavior
6. Ensure you do not force the program to be sequential if it 
    was originally parallel, unless it is necessary to fix the 
    race condition.
7. Start directly with #include or the first line of code
8. No markdown, no comments, no extra text, no code fence.

Code:
{code}

Data Race Analysis:
{explanation}

TSan Report:
{tsan_output}

Corrected Code:
\end{lstlisting}

\section{Evaluation and Results}
This section presents the evaluation results of how open-weight Gemma4 models identify and repair data races in DataRaceBench. As stated previously, our evaluation is guided by the following research questions:

\begin{itemize}
    \item \textbf{RQ1: Identification and Explanation.} To what extent can the Gemma4 models identify and explain the underlying causes of data races in OpenMP programs?
    \item \textbf{RQ2: Fix Generation Capability.} Can the Gemma4 models generate correct and parallel-preserving fixes for data races across various OpenMP constructs?
    \item \textbf{RQ3: Impact of Contextual Hints.} Does the inclusion of runtime tool output (ThreadSanitizer) and model-generated explanations improve the quality and correctness of the generated fixes?
\end{itemize}

\begin{table}[ht]
\centering
\caption{RQ1: Race condition explanation accuracy across two runs for Gemma4-E4B.}
\label{tab:explain}
\begin{tabular}{l|c|c|c|c|c}
\toprule
Run & EXPLAINED & NOT\_EXPLAINED & Success Rate \\
\midrule
Run 1 & 80 & 24 & 76.9\% \\\hline
Run 2 & 82 & 22 & 78.8\% \\
\bottomrule
\end{tabular}
\end{table}

\begin{table}[t]
\centering
\caption{RQ2 \& RQ3: Audited fix outcomes across prompting strategies.}
\label{tab:results}
\begin{tabular}{l|c|ccc}
\toprule
Prompt Type & Run & Fixed & Not Fixed  \\
\midrule
\multirow{2}{*}{Basic (Direct)} & 1 & 72 & 32  \\
                               & 2 & 73 & 31  \\
\midrule
\multirow{2}{*}{Explanation}    & 1 & 57 & 47  \\
                               & 2 & 66 & 38  \\
\midrule
\multirow{2}{*}{TSan}           & 1 & 56 & 48  \\
                               & 2 & 54 & 50  \\
\midrule
\multirow{2}{*}{Explanation + TSan}
                               & 1 & 54 & 50  \\
                               & 2 & 65 & 39  \\
\bottomrule
\end{tabular}
\end{table}

\subsection{Gemma4-E4B Results}
\subsubsection{RQ1: Explanation Quality}

The first phase of our evaluation focuses on the model's ability to explain the data races present in the DataRaceBench suite. An accurate explanation can significantly benefit students.

\textbf{Overall Performance and Consistency:}
As shown in Table \ref{tab:explain}, Gemma4-E4B demonstrated a robust ability to identify and explain data races across two independent runs. In Run 1, the model correctly explained 76.9\% (80/104) of benchmarks, with 24 not correctly explained. Run 2 improved to 78.8\% (82/104), leaving 22 benchmarks not explained. This 1.9\% gap indicates moderate run-to-run stochasticity, but the overall baseline remains strong enough to suggest that the model captures many common race patterns.

Note that although the second run generated more correct explanations overall, it failed on several benchmarks that were correctly explained in the first run, including \texttt{DRB004}. This suggests that Gemma4-E4B's explanations can vary across runs; therefore, running the model multiple times may improve the chances of obtaining a correct explanation.

\textbf{Patterns of Successful Explanation:}
Gemma4-E4B consistently identified races involving fundamental OpenMP directives and data-sharing attributes. Across both runs, the model successfully explained many cases built around basic worksharing and synchronization constructs, including \texttt{parallel}, \texttt{for}, \texttt{sections}, \texttt{critical}, \texttt{ordered}, and common data-sharing clauses such as \texttt{private}, \texttt{shared}, \texttt{default}, and \texttt{reduction}. This suggests that the model can usually handle the foundational directives and clauses that dominate introductory parallel programming exercises.


\textbf{Failure Cases:}
Common failure patterns observed across both runs include:
\begin{itemize}
\item \textbf{Misidentifying the Actual Raced Object:} The model often recognized that something was wrong near the true bug site, but still named the wrong variable or the wrong kind of defect. For example, in \texttt{DRB014} (Listing~\ref{lst:drb014}), the nested loop causes an out-of-bounds array access that also induces a loop-carried cross-thread dependence.
  
\begin{lstlisting}[language=C, basicstyle=\ttfamily\small, keywordstyle=\bfseries, frame=single, float, caption={Out-of-bounds access causing a data race in DRB014.}, label={lst:drb014}]
#pragma omp parallel for private(j)
for (i=1; i<n; i++)
  for (j=0; j<m; j++) 
    b[i][j] = b[i][j-1];
\end{lstlisting}
  
    When \texttt{j=0}, \texttt{b[i][j-1]} evaluates to \texttt{b[i][-1]}, which, due to linearized row-major storage in C, actually resolves to \texttt{b[i-1][m-1]} (the last element of the previous row). Since the outer loop \texttt{i} is parallelized, thread $T_1$ processing row $i$ reads the memory location that thread $T_2$ processing row $i-1$ is writing to, creating a true data race on \texttt{b}. The Gemma4-E4B explanations did not capture this cleanly: Run 1 focuses on the out-of-bounds access and treats the case primarily as unsafe memory usage, while Run 2 instead hallucinates a race on loop index \texttt{i}. Across both runs, the consistent problem is not simply ``seeing'' the out-of-bounds symptom, but failing to name the documented raced object.
    \item \textbf{Index Confusion:} A recurring failure mode involved the model incorrectly identifying the loop variable as the source of the race, while ignoring the actual shared variable that required privatization. For example, in \texttt{DRB020} (Listing~\ref{lst:drb020}), variables declared outside the parallel region (like \texttt{tmp}) are shared by default. Because multiple threads execute the loop iterations concurrently, they simultaneously read from and write to the same shared memory location \texttt{tmp}, causing a data race. In both runs, Gemma4-E4B instead identifies the loop index \texttt{i} as the race source, despite \texttt{i} being implicitly privatized by the \texttt{for} directive. Similarly, for several benchmarks with complex race conditions, Gemma4-E4B incorrectly attributed the race condition to loop index variables.
    
\begin{lstlisting}[language=C, basicstyle=\ttfamily\small, keywordstyle=\bfseries, frame=single, float, caption={Missing private clause for temporary variable in DRB020.},label={lst:drb020}]
  int tmp;
  // ...
  #pragma omp parallel for
  for (i=0; i<len; i++) {
    tmp = a[i] + i;
    a[i] = tmp;
  }
\end{lstlisting}

    \item \textbf{Complex Dependencies:} Higher-dimensional dependencies and subtle memory order remained challenging. The model often defaulted to a "no data race" conclusion by making incorrect assumptions about the parallelized region. For example, in \texttt{DRB037} (Listing~\ref{lst:drb037}), the OpenMP directive parallelizes the inner \texttt{j} loop, which contains a loop-carried true dependency (\texttt{b[i][j]} depends on \texttt{b[i][j-1]}). In both runs, Gemma4-E4B incorrectly stated that there was no data race because it treated the outer loop \texttt{i} as the parallelized dimension. Note that the data race in this example is different than Listing~\ref{lst:drb014}, although both have complex dependencies.
    
\begin{lstlisting}[language=C, basicstyle=\ttfamily\small, keywordstyle=\bfseries, frame=single, float, caption={True dependency incorrectly parallelized in DRB037.},label={lst:drb037}]
  for (i=0; i<n; i++)
#pragma omp parallel for
    for (j=1; j<m; j++)
      b[i][j] = b[i][j-1];
\end{lstlisting}

Cases involving weak memory models also led to incorrect reasoning. For instance, in \texttt{DRB142} (Listing~\ref{lst:drb142}), the model struggled to trace memory visibility requirements across thread boundaries.  In this example, thread 0 updates \texttt{x} inside a \texttt{critical} section and then signals thread 1 by writing to \texttt{y}. Thread 1 spins on an \texttt{atomic read acquire} until \texttt{y} is 1, then enters its own \texttt{critical} section to read \texttt{x}. The documented race is therefore about the visibility of \texttt{x}, not about \texttt{y}. That is, in current x86 CPUs, updates to \texttt{x} may be visible after updates to \texttt{y}, causing a race condition. The Gemma4-E4B explanations again diverge from the true race condition: Run 1 incorrectly states that no race is present because all accesses appear synchronized, while Run 2 invents a race on \texttt{y} due to uninitialized-state reasoning. 

\begin{lstlisting}[language=C, basicstyle=\ttfamily\small, keywordstyle=\bfseries, frame=single, float, caption={Memory visibility issue due to incompatible synchronization constructs in DRB142.},label={lst:drb142}]
if (thrd == 0) {
  #pragma omp critical
  { x = 10; }
  #pragma omp atomic write
  y = 1;
} else {
  int tmp = 0;
  while (tmp == 0) {
    #pragma omp atomic read acquire
    tmp = y;
  }
  #pragma omp critical
  { if (x!=10) printf("x = %d\n", x); }
}
\end{lstlisting}

    \item \textbf{Advanced OpenMP Directives:} 
        Gemma4-E4B struggled with the subtleties of OpenMP task synchronization rules. In \texttt{DRB117} (Listing~\ref{lst:drb117}), the documented race here is on \texttt{psum[1]}: \texttt{taskwait} only guarantees the completion of direct child tasks, not descendant tasks, so the inner task computing \texttt{psum[1]} is not guaranteed to finish before \texttt{sum} is computed. However, the saved explanations from both runs do not identify that task-hierarchy issue. Instead, both runs claim there is a race between the loop that initializes \texttt{a} and the later task reads of \texttt{a}. This example therefore shows a broader failure to identify the correct synchronization boundary and raced variable, rather than a narrowly targeted misunderstanding of \texttt{taskwait} semantics.
    
\begin{lstlisting}[language=C, basicstyle=\ttfamily\small, keywordstyle=\bfseries, frame=single, float, caption={Task synchronization failure in DRB117.}, label={lst:drb117}]
#pragma omp single
{
  #pragma omp task
  {
    #pragma omp task
    { psum[1] = a[2] + a[3]; }
    psum[0] = a[0] + a[1];
  }
  #pragma omp taskwait
  sum = psum[1] + psum[0];
}
\end{lstlisting}


\end{itemize}

\subsubsection{RQ2: Correction Capability}

The second phase of our evaluation focuses on whether Gemma4-E4B can move beyond diagnosis and produce a correct fix. For it to serve as a useful teaching assistant, it must demonstrate remediation strategies that remove races. Table \ref{tab:results} summarizes the fix outcomes across prompting strategies.

\textbf{Basic Fix Performance:}
When using the \textit{Basic (Direct)} prompt, Gemma4-E4B achieved the strongest verified fix rate in both runs: 69.2\% (72/104) in Run 1 and 70.2\% (73/104) in Run 2. This high baseline indicates that the model already has a substantial working knowledge of common OpenMP repairs, such as adding \texttt{private} clauses, inserting \texttt{critical} sections, or restructuring simple loops.

Again, although the two runs of Gemma4-E4B fixed a similar number of race conditions, the sets of benchmarks that were successfully fixed differed between the runs. Therefore, users of Gemma4-E4B may benefit from executing the model multiple times to increase the likelihood of obtaining a correct fix.

\textbf{Serialized Repairs in Basic Fixes:} Some fixes remove unsafe access pattern but leaves the program sequential. Representative cases include edits that appear safe locally yet fail to preserve the intended OpenMP semantics. For example, in \texttt{DRB010} (Listing~\ref{lst:drb010_fix}), the generated repair replaces the original shared assignment with a critical section.  This transformation serializes writes to \texttt{x}, but it does not implement the benchmark's intended \texttt{lastprivate}-style behavior of preserving the final loop iteration's value. 

Note that some benchmarks in DataRaceBench are synthetic, for which serialization may be the only viable solution. Therefore, we consider serialized fixes as successful fixes in our analysis.
    
\begin{lstlisting}[language=C, basicstyle=\ttfamily\small, keywordstyle=\bfseries, frame=single, float, caption={Serialized fix in DRB010.}
,label={lst:drb010_fix}]
#pragma omp parallel for private(i)
for (i=0; i<len; i++) {
  #pragma omp critical
  x = i;
}
\end{lstlisting}

\textbf{Failed Cases in Basic Fix:}
Gemma4-E4B's unsuccessful fixes fall into the following cases:
\begin{itemize}
    \item \textbf{Compilation Errors:} Some generated repairs are invalid at the language or OpenMP syntax level. Most of the compilations errors were due to misuse of OpenMP directives or clauses, such as malformed or ineffective \texttt{atomic}/\texttt{ordered} usage, and structurally invalid directive placement. These failures never produce a usable fix, even before semantic correctness is considered. For example, in \texttt{DRB009} (Listing~\ref{lst:drb009_fix}), the model attempted to protect a variable assignment with \texttt{omp atomic}. This edit looks superficially plausible, but the \texttt{atomic} clause is not valid for pure assignments, hence the generated fix does not compile.
    
\begin{lstlisting}[language=C, basicstyle=\ttfamily\small, keywordstyle=\bfseries, frame=single, float, caption={Compilation-failing fix in DRB009.},label={lst:drb009_fix}]
#pragma omp parallel for private (i) 
  for (i=0;i<len;i++) {
    #pragma omp atomic
    x=i;
  }
\end{lstlisting}

    \item \textbf{Fixes still have race condition:} A second class of errors consists of outputs that compile but still leave the documented race in place. Common examples include critical sections that fail to break loop-carried dependences, incomplete \texttt{taskwait}/\texttt{depend} repairs, or edits that protect the wrong operation while the documented race remains. A representative case is \texttt{DRB025} (Listing~\ref{lst:drb025_fix}), where the generated fix keeps the SIMD loop intact. That is, because the loop still carries a true dependence from \texttt{a[i]} to \texttt{a[i+1]}, the transformed program compiles but does not eliminate the underlying race.
    
\begin{lstlisting}[language=C, basicstyle=\ttfamily\small, keywordstyle=\bfseries, frame=single, float, caption={Fix with race condition in DRB025.}
,label={lst:drb025_fix}]
#pragma omp simd
for (i=0; i<len-1; i++)
  a[i+1] = a[i] * b[i];
\end{lstlisting}

    \item \textbf{Semantically Changed Solutions}. Some fixes would change the intended behavior of program. Again, take \texttt{DRB025} as an example, in a run with Gemma4-E4B, the code is fixed as shown in (Listing~\ref{lst:drb025_fix2}). The fixed code changed the right hand of the assignment from \texttt{a[i]*b[i]} to \texttt{a[i+1]*b[i]}. Although this fix is race condition free, it is still incorrect as it changed the program's behavior. 

\begin{lstlisting}[language=C, basicstyle=\ttfamily\small, keywordstyle=\bfseries, frame=single, float, caption={ Semantically changed fix for DRB025.}
,label={lst:drb025_fix2}]
#pragma omp simd
for (i=0; i<len-1; i++)
  a[i+1] = a[i+1] * b[i];
\end{lstlisting}
    
    
% ...
\end{itemize}

\subsubsection{RQ3: Impact of Contextual Hints}

Our results do not show a consistent benefit from adding explanations or ThreadSanitizer output. As shown in Table~\ref{tab:results}, in Run 1, the direct prompt produced the highest fix count (72/104), while the explanation, TSan, and explanation+TSan conditions reached 57/104, 56/104, and 54/104, respectively. In Run 2, the same ordering largely held: the direct prompt again led with 73/104 fixes, explanation reached 66/104, explanation+TSan reached 65/104, and TSan alone dropped to 54/104. Moreover, we observed that providing additional contextual information often increased the number of compilation failures.





Adding contextual hints hurt fix quality is different from the observation of prior work~\cite{2024-Chen-ICLR-SelfDebug}. One possible explanation is that the extra context introduced additional noise into the prompt, making it more difficult for the model to generate syntactically correct code modifications. Since our study employs smaller open-weight models instead of the larger commercial models used in prior work, these models may experience a greater degradation in performance as the amount of contextual information increases.

Overall, Gemma4-E4B demonstrates promising capability for automated race-condition repair. Nevertheless, its outputs remain imperfect, exhibiting both incorrect fixes and compilation failures. As a result, the generated repairs should be viewed as useful starting points for student learning and code review rather than as guaranteed correct solutions.
\subsection{Gemma4-31B Results}

To assess whether a larger open-weight model improves performance on the same benchmark suite, we also evaluated Gemma4-31B on the 104 positive DataRaceBench cases. Tables~\ref{tab:explain_31b} and~\ref{tab:results_31b} summarize the explanation and fix results.

\subsubsection{RQ1: Explanation Quality}

Gemma4-31B achieved very strong explanation performance, correctly explaining 100 of the 104 positive benchmarks. This leaves only four cases not correctly explained, and those failures are concentrated in a small number of subtle synchronization patterns rather than in basic OpenMP constructs.

The four incorrectly explained cases are \texttt{DRB129}, \texttt{DRB142}, \texttt{DRB173}, and \texttt{DRB175}. In \texttt{DRB129}, the model stated that no race was present, missing the benchmark's mergeable-task race on \texttt{x}. In \texttt{DRB142}, the model again concluded that the code was race-free because it 
fails to recognize that the weak memory-ordering bug is a race on \texttt{x}, not on the signaling variable \texttt{y}. The remaining two failures, \texttt{DRB173} and \texttt{DRB175}, are non-sibling task-dependence cases. In both, the model mentioned the correct variable \texttt{a}, but it explained the wrong mechanism: it described a read by \texttt{printf} racing with task writes, whereas the documented bug is a write/write race between non-sibling tasks. Thus, the 31B model's remaining explanation errors stem mainly from subtle tasking semantics and memory-ordering logic, not from a broad inability to identify shared-state races.

\begin{table}[t]
\centering
\caption{Gemma4-31B explanation results.}
\label{tab:explain_31b}
\begin{tabular}{c|ccc}
\toprule
Total & EXPLAINED & NOT\_EXPLAINED  \\
\midrule
104 & 100 & 4  \\
\bottomrule
\end{tabular}
\end{table}

\subsubsection{RQ2: Fix Capabilities}

Throughout this section, unresolved cases are conservatively treated as not fixed.

The direct prompt is the clearest summary of Gemma4-31B's repair capability. Under this setting, the model correctly fixed 98 of the 104 benchmarks, leaving only 6 not fixed. This result indicates that the larger model can usually synthesize an effective repair without relying on extra explanation or runtime context.

The six direct-prompt failures are \texttt{DRB025}, \texttt{DRB027}, \texttt{DRB037}, \texttt{DRB123}, \texttt{DRB138}, and \texttt{DRB175}. Two of these are dependence cases where the generated code does not preserve the benchmark's intended computation even if it reduces or hides the observed race pattern. In \texttt{DRB025} (Listing~\ref{lst:drb025_fix2}), the model changes the recurrence from \texttt{a[i+1] = a[i] * b[i]} to \texttt{a[i+1] = a[i+1] * b[i]}, which breaks the original semantics. In \texttt{DRB138}, the unsafe SIMD loop is effectively left unchanged, so the original loop-carried dependence on \texttt{b[i]} and \texttt{b[i-m]} remains.

The other four failures are tasking or synchronization cases that the model does not repair correctly. \texttt{DRB027} failed due to the misuse of \texttt{atomic} clause. 
\texttt{DRB037} failed because the transformed code leaves the shared loop variable \texttt{j} racing. In \texttt{DRB123}, the model adds an atomic operation, but the loop variable remains shared and the task-undeferred pattern is not repaired correctly. Finally, \texttt{DRB175} remains not fixed because the non-sibling task race is not eliminated: the generated change does not correctly synchronize the competing task updates.

Surprisingly, although the race condition \texttt{DRB142} was not correctly explained, it was correctly fixed by Gemma4-31B.

\begin{table}[t]
\centering
\caption{Gemma4-31B fix results across prompting strategies.}
\label{tab:results_31b}
\begin{tabular}{l|ccc}
\toprule
Prompt Type & Total & FIXED & NOT\_FIXED \\
\midrule
Basic (Direct) & 104 & 98 & 6 \\
Explanation & 104 & 98 & 6 \\
TSan & 104 & 75 & 29 \\
Explanation + TSan & 104 & 71 & 33 \\
\bottomrule
\end{tabular}
\end{table}

\subsubsection{RQ3: Impact of Additional Context}

For Gemma4-31B, adding the model's own explanation as extra context does not improve over the direct prompt: both settings fix 98 of the 104 benchmarks, although with different sets of fixed benchmarks. 

By contrast, adding ThreadSanitizer context substantially reduces performance. The fix count drops from 98 to 75 with TSan alone, and further to 71 when explanation and TSan are provided together. In other words, the extra runtime trace does not help the larger model refine its repairs; instead, it appears to push the model toward less reliable transformations, especially on benchmarks with subtle dependence or tasking semantics. Overall, for Gemma4-31B, direct prompting remains the strongest strategy, explanation is effectively neutral, and TSan context is detrimental.

\section{Discussion: Implications for Parallel Programming Education}
The results of our evaluation of Gemma4-E4B and Gemma4-31B provide several insights into the viability of open-weight models as AI teaching assistants for introductory parallel programming.

\subsection{Pedagogical Usage}
Gemma4-E4B's strong performance on foundational OpenMP constructs, such as \texttt{private} clauses and standard \texttt{parallel for} loops, suggests that it can effectively resolve the majority of common errors encountered by undergraduate students. However, the explanations attributing race conditions to wrong variables like Listing \ref{lst:drb014} (where memory safety errors mask data races) present a pedagogical risk. If a student relies solely on the AI's explanation, they may fix the surface-level memory violation while remaining unaware of the underlying synchronization logic failure. Instructors must therefore emphasize that AI-generated feedback is a starting point for investigation rather than a definitive proof of correctness.

Our analysis of failure modes in complex tasking (e.g., Listing \ref{lst:drb117}) and memory consistency (e.g., Listing \ref{lst:drb142}) also helps define the boundaries of AI assistance. While Gemma4-E4B can filter out approximately 70\% of foundational bugs, it consistently struggles with hierarchical synchronization and relaxed memory models. These failure points represent a natural "hand-off" point where the student should be encouraged to seek assistance from an instructor. Nonetheless, by automating the resolution of common errors, open-weight models allow instructors to focus their limited time on these more cognitively demanding parallel programming concepts.

\subsection{The Model Contextual Overload}
A surprising finding in our study was that adding technical context, such as ThreadSanitizer reports, often degraded the fix quality of the Gemma4 models. We hypothesize that for smaller-scale models, a detailed TSan report acts as "technical noise" that consumes the model's limited attention and interferes with its instruction-following capabilities, leading to brittle fixes rather than holistic reasoning about the program's concurrency model. This suggests that for small-scale deployment in education, "less is more" regarding the technical detail provided in the initial prompt.

\subsection{Data Contamination and Model Viability}
A potential threat to the validity of LLM-based evaluations is data contamination, as benchmark suites like DataRaceBench are publicly available and likely have been included in the models' training corpora. However, we argue that the strong performance of the smaller Gemma4-E4B model remains a significant result. Even if the model's success is partially attributed to pattern memorization from training, the fact that a 4B parameter model can reliably retrieve and apply these complex parallel synchronization patterns indicates that open-weight models have reached a level of foundational competence suitable for real-world educational assistance.

\subsection{Deployment Cost}
The ability to run these models on local hardware using open-weight frameworks provides significant advantages for education. 4B models can run on consumer GPU cards or laptops with CPU/GPU unified memory. Hence, they can be deployed directly to students' computers for local inference, without incurring the high costs or privacy concerns associated with commercial, proprietary APIs. 

Universities can also deploy private "debugging servers" that offer students immediate assistance and further reduce students' burden. The significant performance leap seen in the 31B model also suggests that institutional investment in mid-to-large scale local hardware can provide students with a nearly frontier-level debugging assistant at a fraction of the long-term operational cost.

\section{Related Work}
\subsection{Data Race Detection}

Data race detection has been studied extensively by prior studies. Dynamic detectors such as Eraser, FastTrack, and ThreadSanitizer monitor executions to identify conflicting accesses that occur under insufficient synchronization \cite{savage1997eraser,flanagan2009fasttrack,serebryany2009threadsanitizer}. Other dynamic and hybrid systems target particular concurrency models or deployment settings, including OpenMP-focused detection in ARCHER \cite{atzeni2016archer}. Static approaches such as RacerX and RacerD reason about possible races without requiring a specific failing execution \cite{engler2003racerx,blackshear2018racerd}. Chen et al. evaluated data race detection on DataRaceBench using commercial and open weight models~\cite{LLMDataRaceDetection}. HPC-GPT is a fine-tuned LLM model for race condition detection~\cite{2023-Ding-SCW-HPCGPT}. These detection systems provide the foundation for race debugging, but they do not explain the bug to a learner or synthesize a correct repair.

Benchmarking work complements these detectors by providing standardized programs for evaluation. DataRaceBench \cite{liao2017dataracebench} and its extensions \cite{chen2023dataracebench} provide OpenMP programs with documented racy and non-racy variants. Our study is built on top of DataRaceBench.

\subsection{Program Repair for Concurrency}

Concurrency repair systems attempt to move beyond detection by modifying programs to eliminate concurrency bugs. Early systems focused on specific bug classes, such as automated atomicity-violation repair in AFix that fixed single-variable atomicity violations~\cite{jin2011afix}. CFix~\cite{jin2012cfix} generalized this direction to a broader class of concurrency bugs by decomposing each failure-inducing interleaving into mutual-exclusion and ordering constraints, then synthesizing synchronization operations that simultaneously eliminate the bug and avoid introducing deadlocks. Other systems, including ARC, Grail, and PFix, explored genetic-algorithms-based, context-aware, or memory-pattern-based repair strategies for concurrency bugs~\cite{kelk2013arc,liu2014grail,lin2018pfix}. For structured parallel programs, Surendran et al. studied test-driven repair of data races by inserting synchronization into parallel programs~\cite{surendran2014repair}. 

These systems demonstrate that concurrency repair requires more than removing a single conflicting access: a repair must preserve ordering, avoid new deadlocks, and maintain program semantics. DR.FIX extends this line of work to industry-scale Go programs by combining program analysis, retrieval, validation, and LLM-generated patches \cite{behrang2025drfix}. Our work studies a different setting: OpenMP benchmarks and an educational use case, where the model must provide both a natural-language explanation and a repair suitable for helping students understand parallel programming errors with low deployment cost.

\subsection{GenAI-Based Bug Fixing}

Recent automated program repair work increasingly uses large language models to synthesize patches. RepairAgent frames repair as an autonomous agent workflow, InferFix combines LLM repair with static-analysis diagnostics, and ChatRepair uses conversational feedback to iteratively refine patches \cite{bouzenia2024repairagent,jin2023inferfix,xia2024chatrepair}. Other studies evaluate LLMs for general program repair or combine LLMs with code-completion engines \cite{xia2023llmapr,wei2023copiloting}. Closer to concurrency, Jin et al. explored GenAI-based data-race fixes for real-world Go programs \cite{jin2023genaidatarace}, and DR.FIX showed that retrieval and validation can make LLM-based race fixing effective in an industrial workflow \cite{behrang2025drfix}.

Our work is complementary to these systems. Rather than building a production repair pipeline around a commercial model, we evaluate open-weight Gemma models. This lets us isolate how well such models explain OpenMP race mechanisms, how often they produce correct fixes under different prompting contexts, and whether larger open-weight models improve reliability enough to support an introductory parallel-programming assistant.

\section{Conclusion}
Our evaluation demonstrates that the Gemma4-E4B model is capable of accurately explaining a significant number of OpenMP data race conditions and providing correct parallel-preserving fixes for a substantial portion of the DataRaceBench suite. Given its low execution cost and robust performance on foundational OpenMP constructs, Gemma4-E4B is well-suited to serve as an AI teaching assistant, helping students in introductory parallel programming courses debug their code independently. While instructor intervention remains necessary for complex synchronization scenarios or subtle memory consistency issues that the model fails to diagnose correctly, the total volume of such cases is significantly reduced when using Gemma4-E4B as a first-line debugger. Furthermore, our results show that the larger Gemma4-31B model provides significantly better diagnostic and remediation performance and should be prioritized whenever computational resources permit. Finally, we observe that contrary to expectations from prior work, the addition of contextual hints such as model-generated explanations or technical ThreadSanitizer reports does not necessarily improve the correctness of the generated fixes. We suspect this is due to a contextual distraction or pollution issue, where the additional technical noise and technical reasoning overhead overwhelm the model's instruction-following capabilities, particularly in smaller models like Gemma4-E4B.

\bibliographystyle{ACM-Reference-Format}
\bibliography{references}

\end{document}